\documentclass[12pt,a4paper]{article}   
\usepackage[dvips]{graphicx}
\usepackage{amsmath,amsfonts,amssymb}

\newtheorem{thm}{Theorem}

\begin{document}
\title{
Functional relations in Stokes multipliers 
\\
- Fun with $x^6 + \alpha x^2$ potential-
}
\author{ J. Suzuki \thanks{e-mail: sjsuzuk@ipc.shizuoka.ac.jp}\\
           \parbox{0.9\textwidth}{
        {\em
        \begin{center}
         Department of Physics, Faculty of Science, \\
	Shizuoka University, \\
	 836 Ohya, Shizuoka 422, \\
	 Japan
        \end{center}
        }}
       }
\maketitle
\begin{center}
-Dedicated to Professor R. J. Baxter on the occasion of his 60-th birthday-
\end{center}
\begin{abstract}
We consider eigenvalue problems in quantum mechanics
in one dimension.
Hamiltonians  contain  a class of double well potential terms,
$x^6 + \alpha x^2$ , for example .
The space coordinate is continued to a complex plane and 
the connection problem of fundamental system of solutions is considered.
A hidden $U_q(\widehat{gl}(2|1))$ structure arises
in "fusion relations" of Stokes multipliers.
With this observation, we derive coupled nonlinear integral
equations which characterize the spectral properties  of
both $\pm \alpha$ potentials simultaneously.
\end{abstract}


\clearpage
\section{Introduction}

The eigenvalue problem of a one-body 
1D Schr{\"o}dinger operator
is the most fundamental subject in quantum mechanics.
Still, it provides vivid materials of research.

Besides a few exceptions where eigenvalues and wavefunctions 
are obtainable explicitly, one may employ several tools for analysis,
e.g., the perturbation theory, the variational approach and so on.
Among them, the exact WKB method \cite{V1}-\cite{Takei}
is unique in the sense that it
provides non-perturbative 
information on the analytical structure of wavefunctions 
and spectral properties.
We analytically continue $x$, original coordinate variable, 
 to a complex number.
The whole complex plane is divided into several sectors.
In each sectors, there are two linearly independent solutions
, as Schr{\"o}dinger operator is the 2nd order differential
operator. 
They are referred to as  the fundamental system of solutions
(FSS) in the sector \cite{FBOOK}.
The relations among FSS in different sectors are central
issue in the connection problem.
The importance of the problem 
and consequently Stokes multipliers,
in the WKB problem has been deeply recognized 
and emphasized in early '80, especially in \cite{V1}.

Recently, a remarkable link has been established among
the spectral determinants of a 1D Schr{\"o}dinger operator
associated with the anharmonic oscillator, 
transfer matrices and ${\bf Q}$ operators in CFT  
possessing $U_q(\widehat{sl}(2))$
\cite{DT1}-\cite{JS}.
Here the spectral determinants imply 
$D(E)=\prod_{E_j \in {\rm eigenvalue}} ( 1- \frac{E}{E_j})$ 
and its generalizations.
A curious interplay between $D(E)$ and
 generalized Stokes multipliers is also found \cite{V1, DT2}.
In view of solvable models, a striking fact is that 
they share the same functional relations with transfer matrices 
in the fusion hierarchy  possessing $U_q(\widehat{sl}(2))$ 
\cite{DT2, Sbook, Baxbook,BLZ1}.
This allows for applications of the strong machineries in the study of
solvable models \cite{Baxbook}-\cite{SuzSpinS} 
to the studies of Stokes multipliers, spectral
determinants and so on.
Several results have been explicitly obtained 
for the anharmonic oscillator problem,
and  are extended to higher differential analogues \cite{DT3, JSAN}.

In this note,  we consider an anharmonic oscillator  
perturbed by a lower power potential term.
It belongs to a class of potentials discussed 
generally in \cite{V6,V7} with the 
exact resolution method.
To be precise, we consider the eigenvalue problem,
\begin{equation}
{\cal H}(x,\alpha) \Psi_k(x)=
\Bigl( -\frac{d^2}{dx^2} + x^{2M} + \alpha  x^{M-1} \Bigr ) \Psi_k(x)= E_k \Psi_k(x).
\label{hamil}
\end{equation}
Throughout this report we set $\hbar=1$ and $M >1$.

The spectral problem concerning this Hamiltonian turns out to be 
in a category to which one can apply the tools in
solvable models.

The sign of $\alpha$ seems to be crucial if one
considers the operator (\ref{hamil}) on the real axis.

\begin{equation*}
\boxed{\rm FIG. 1}
\end{equation*}

We will not expect much difference from the "pure" anharmonic
oscillator when $\alpha>0$, while we do expect change for $\alpha<0$
as the potential develops the double well.

It will be shown, however, that the negative $\alpha$ 
and the positive $\alpha$ problems 
are not separable when we discuss the global connection problem.
Roughly speaking, 
the negative $\alpha$ problem is coupled to 
the positive $\alpha$ problem by crossing a border line of
neighboring sectors and vice versa.
See \S 2 for precise arguments.
It may be then reasonable to consider a two-fold connection problem
(crossing two adjacent lines) , or more generally, 
relations between sectors
separated by even multiples of border lines.
Some of the Stokes multipliers, in the generalized connection problem, 
possess expressions corresponding to the eigenvalues of
the (fusion) transfer matrices of the 3 state Perk-Schulz (PS) model 
\cite{PerkSchulz81, Schulz81,Schulz83}
of which underlying symmetry is  $U_q(\widehat{gl}(2|1))$.
Others can not be directly equated with the (fusion) transfer matrices
but  have relations with the the 3 state PS model as well.
Thus  we conclude that the perturbation $\alpha  x^{M-1} $
breaks the $U_q(\widehat{sl}(2))$ symmetry of the "pure"
anharmonic oscillator but it brings the new symmetry 
$U_q(\widehat{gl}(2|1))$.
The deformation parameter $q$ is related to the exponent
of the perturbation by $q=\exp(i\frac{\pi}{M+1})$.
Through these findings, we can  
derive the nonlinear integral equations (NLIE) which characterize
the energy levels of both the negative $\alpha$ problem and 
the positive $\alpha$ problem simultaneously.

The paper is organized as follows.
In the next section, we will explore symmetries of solutions
to (\ref{hamil}). The precise definition of sectors is given.
The connection problem is addressed in section 3.
Certain components in fusion Stokes matrices are identified with
eigenvalues of fusion transfer matrices associated to $U_q(\widehat{gl}(2|1))$.
The spectral determinant is  explicitly parameterized by 
FSS in a sector.
The coupled NLIE are  then derived in section4,
which determine energy levels.
We will perform analytical and numerical checks on the consistency
of our result in section 5.
Section 6 is devoted to summary and discussions on open problems.


\section{Asymptotic Expansion and Symmetry of solutions}

Let $\phi(x,\alpha,E) $ be an entire function of $(x,\alpha, E)$
and a  solution to
${\cal H}(x,\alpha) \phi(x,\alpha,E) = E \phi(x,\alpha,E) $.

The solution, 
which decays exponentially at $x \rightarrow \infty$, is of primary interest.
By employing the argument in \cite{HS}, we immediately find
its asymptotic behavior,

\begin{eqnarray}
\phi(x, \alpha, E) & \sim&  x^{-M/2 -\alpha/2} 
 \exp(-\frac{x^{M+1}}{M+1}), \label{asymptotics} \\
\partial_x \phi(x, \alpha, E) & \sim&  x^{M/2 -\alpha/2} 
 \exp(-\frac{x^{M+1}}{M+1}). \label{asymptotics2}
\end{eqnarray}
The validity of the above expansion is not restricted to
the real axis, but extends to
the wedge, $|{\rm arg} x |< \frac{3 \pi}{2M+2}$ \cite{ Sbook, HS}.

The second order linear differential equation admits another
independent solution.
To specify it, or to deal with the global problem, it is
convenient to extend  $x$ to the complex plane
as mentioned in introduction.
Then, as in the case of $\alpha=0$, the solution exhibits 
a symmetry  by rotating the complex $x$ plane by
a specific angle.

The direct calculation proves the following.
\begin{thm}
Let $\phi(x,\alpha,E)$ be the above solution and $q=\exp(i\frac{\pi}{M+1})$.
Then $\phi(q^{-1} x, q^{M+1} \alpha, q^2 E)$ is also the solution
to the differential equation, ${\cal H}(x,\alpha) \phi= E \phi$.
\end{thm}

This is the desired second solution  which
grows exponentially on the positive real axis:
$
x^{-M/2 +\alpha/2} 
 \exp(\frac{x^{M+1}}{M+1})$ for $ x \rightarrow \infty$
We note that $q^{M+1} \alpha=-\alpha$.
This deserves an attention.
As mentioned in introduction, the potential assumes 
the completely different
structure for $\alpha$ positive and $\alpha$ negative on the real axis. 
The rotation in the complex $x$ plane by angle 
$\frac{\pi}{M+1}$, however, couples these two problems.
Thus we shall treat the Hamiltonians with $\pm \alpha$
{\it simultaneously}.
Similar pairing of differential equations
is found for the positive and the negative
angular momentum terms in a class of 
3rd order differential equations\cite{DT3}.

This observation is crucial in our approach and
can be generalized further.
To state it, we prepare some notations.

Hereafter  $\alpha$ always takes a non-negative real value.
By ${\cal H}^{(\epsilon)}(x,\alpha)$, we mean
the Schr{\"o}dinger operator,
$$
 -\frac{d^2}{dx^2} + x^{2M} +\epsilon \alpha x^{M-1}
$$
where $\epsilon=\pm 1$.

Let ${\cal S}_k$ be a sector in the plane satisfying
$$
|{\rm arg} x -\frac{k \pi}{M+1}  | \le \frac{\pi}{2M+2}.  
$$

\begin{equation*}
\boxed{\rm FIG. 2}
\end{equation*}

The FSS depends on the sector.
We  conveniently define
$$
y_j^{(\epsilon)}:= 
\frac{q^{j/2-\epsilon \alpha/2}}{\sqrt{2 i}}
 \phi(xq^{-j}, \epsilon \alpha, q^{2j} E).
$$

\begin{thm} 
For the ${\cal H}^{(\epsilon)}(x,\alpha)$, the FSS
in the sector ${\cal S}_{j}$ is given by
$(y_j^{(\epsilon_j)}, y_{j+1}^{(\epsilon_{j+1})})$ where
$\epsilon_j=\epsilon (-1)^{j}$.
\end{thm}

For $\alpha=0$ case, this has been argued in \cite{Sbook, DT2}.
It is easily checked that $y_j^{(\epsilon_j)}$ is the 
sub-dominant solution in ${\cal S}_j$; it tends to zero
as $x$ tends to infinity along in any direction in the sector.

In the next section, we consider the global connection problem
of these FSS in the complex $x$ plane.
%
%

\section{Fusion Stokes multipliers, $U_q(\widehat{gl}(2|1))$ structure
 and Spectral Determinants}

We introduce the Wronskian matrix 
\begin{equation}
\Phi_j^{(\epsilon)}(x):=
\begin{pmatrix}
y_j^{(\epsilon)},            &  y_{j+1}^{(-\epsilon)}   \\
\partial_x y_j^{(\epsilon)},   & \partial_x y_{j+1}^{(-\epsilon)} \\
\end{pmatrix}.
\end{equation}
and the Wronskian $W_k^{(\epsilon)}:
={\hbox {\rm det}} \Phi_k^{(\epsilon)}(x)$.

The linear dependence of the solution can be easily verified
by evaluating the Wronskian at ${\cal S}_{j+1/2}$ 
using the asymptotic expansions (\ref{asymptotics}) and
(\ref{asymptotics2}) .
The present normalization yields $W_k^{(\epsilon)}=1$.

 Let  ${\cal M}_{j,1}^{(\epsilon)}$ 
be the Stokes matrix connecting the Wronskian matrices 
$\Phi_j^{(\epsilon)}(x)$ and $\Phi_{j+1}^{(-\epsilon)}(x)$,
\begin{equation}
\Phi_{j}^{(\epsilon)}(x) =
 \Phi_{j+1}^{(-\epsilon)}(x){\cal M}_{j,1}^{(\epsilon)}.
 \label{single}
\end{equation}
It permits an explicit parameterization

\begin{equation}
{\cal M}_{j,1}^{(\epsilon)}:=
\begin{pmatrix}
\tau^{(\epsilon)}_j,            &  1  \\
-1,   &                            0 \\
\end{pmatrix}, 
\end{equation}
where $ \tau^{(\epsilon)}_j $ is referred to as the Stokes multiplier.
We have two remarks.
First, the (1,1) element is the function of $\alpha$ and $q^{2 j}E$.
We omit the dependency on $\alpha$. The dependency on $E$ is indicated 
by the index $j$.
Second, the (2,1) element,$-1$, 
is a consequence of the present normalization
of the Wronskian. 

To be more specific, we consider the operator ${\cal H}^{(+)}(x,\alpha)$
and start from the positive real axis (or more generally ${\cal S}_0$). 
The initial FSS is $(y_0^{(+)}, y_1^{(-)})$.

\begin{equation*}
\boxed{\rm FIG. 3}
\end{equation*}

A linear relation follows from (\ref{single}) 
 between this FSS and $(y_1^{(-)}, y_2^{(+)})$,
the FSS at the neighboring sector ${\cal S}_1$, 
\begin{equation}
y_0^{(+)}= \tau^{(+)}_0 y_1^{(-)} -y_2^{(+)}. 
\label{taupy}
\end{equation}
Similarly, the FSS in ${\cal S}_2$ is linked to the FSS in ${\cal S}_1$
by
\begin{equation}
y_1^{(-)}=\tau^{(-)}_0 y_2^{(+)} -y_3^{(-)}. 
\label{taumy}
\end{equation}

Judging from the upper indices  which indicate the corresponding signs of $\alpha$,
it may be natural to introduce a generalized Stokes matrix 
${\cal M}_{0,2}^{(+)}$ connecting
 FSS $(y_0^{(+)}, y_1^{(-)})$ and $(y_2^{(+)}, y_3^{(-)})$.
It is simply obtained by the matrix multiplication, 

\begin{equation}
{\cal M}_{0,2}^{(+)}= {\cal M}_{1,1}^{(-)} {\cal M}_{0,1}^{(+)}=
\begin{pmatrix}
\tau^{(-)}_1 \tau^{(+)}_0-1,            &  \tau^{(-)}_1  \\
-\tau^{(+)}_0,   &                            -1 \\
\end{pmatrix}.
\label{fusion2}
\end{equation}

Equations (\ref{taupy}) and (\ref{taumy}) yield $\tau$'s in terms of $y$'s.
The (1,1) component in (\ref{fusion2}), hereafter denoted by
$T_{1,1}(E)$,  is then represented 
 in terms of $y$ as, 
\begin{equation}
T_{1,1}(E)=
\tau^{(-)}_1 \tau^{(+)}_0-1=
\frac{y_0^{(+)}}{y_2^{(+)}}+
\frac{y_0^{(+)}y_3^{(-)}}{y_2^{(+)}y_1^{(-)}} +
\frac{y_3^{(-)}}{y_1^{(-)}}.
\label{pstransfer}
\end{equation}
The dependence of $E$ in the rhs is implicitly
indicated by indices of $y$.
We will comment on this representation in terms of a solvable model later.

There is another expression using both $y$'s and $\partial y$'s.
This form is of practical use in the following generalization.
By applying the Cramer formula to (\ref{single}),
we immediately obtain

\begin{equation}
\tau^{(\epsilon)}_j = 
\begin{vmatrix}
y_j^{(\epsilon)},            &  y_{j+2}^{(\epsilon)}  \\
\partial_x y_j^{(\epsilon)},   &  \partial_x y_{j+2}^{(\epsilon)} \\
\end{vmatrix}.
\label{qwronskian}
\end{equation}
Note that we use the fact that the Wronskian is normalized to be unity.
The (1,1) entries in (\ref{fusion2}) is then given by
$$
\begin{vmatrix}
y_0^{(+)},            &  y_{3}^{(-)}  \\
\partial_x y_0^{(+)},   &  \partial_x  y_{3}^{(-)}  \\
\end{vmatrix}.
$$

One can further generalize the above result.
Naturally, the "fusion" Stokes matrix ${\cal M}_{j,2 k}^{(+)}$ 
is defined  which 
relates FSS of ${\cal S}_j$ to ${\cal S}_{j+2k}$.
Explicitly, it is given by

\begin{equation}
{\cal M}_{0,2k}^{(+)}=
\begin{pmatrix}
\begin{vmatrix}
y_0^{(+)},            &  y_{2k+1}^{(-)}  \\
\partial_x y_0^{(+)},   &  \partial_x  y_{2k+1}^{(-)}  \\
\end{vmatrix}, &
\begin{vmatrix}
y_1^{(-)},            &  y_{2k+1}^{(-)}  \\
\partial_x y_1^{(-)},   &  \partial_x  y_{2k+1}^{(-)}  \\
\end{vmatrix}  \\
-\begin{vmatrix}
y_0^{(+)},            &  y_{2k}^{(+)}  \\
\partial_x y_0^{(+)},   &  \partial_x  y_{2k}^{(+)}  \\
\end{vmatrix},   &  
-\begin{vmatrix}
y_1^{(-)},            &  y_{2k}^{(+)}  \\
\partial_x y_1^{(-)},   &  \partial_x  y_{2k}^{(+)}  \\
\end{vmatrix} \\
\end{pmatrix},
\label{fusion2d}
\end{equation}
for $j=0$.
We can prove the above using the induction on $k$ most easily.
Similar formula holds for ${\cal M}_{0,2 k}^{(-)}$ by replacing all
upper indices $+ \leftrightarrow -$.

We are now ready to relate an entry in a fusion Stokes matrix to
the spectral determinant.
Hereafter we assume $M=2m-1$.
It follows from the above argument that
\begin{equation}
\Phi_{2m }^{(+)}= \Phi_{0}^{(+)} ({\cal M}_{0,2 m}^{(+)})^{-1}.
\label{posneg}
\end{equation}

\begin{equation*}
\boxed{\rm FIG. 4}
\end{equation*}

 $y_{0}^{(+)}$( $y_{2m}^{(+)}$) stands for 
the subdominant solution on the positive (negative)
real axis. 
They tend to zero asymptotically in their proper region, being appropriate
basis for the eigenfunction.  
The eq.(\ref{posneg}) tells, however, that $y_{2 m }^{(+)}$ is combined to
 {\it both} $y_{0}^{(+)}$ {\it and} $y_{1}^{(-)}$ by rotating the complex
 plane by $-\pi$, 
 
\begin{eqnarray*}
y_{2m}^{(+)} &=& 
(c_1 y_0^{(+)} +c_2 y_1^{(-)} )/{\rm det} {\cal M}_{0,2 m}^{(+)}\\
c_1&=& ({\cal M}_{0,2 m}^{(+)})_{1,1}, \quad 
c_2= -({\cal M}_{0,2 m}^{(+)})_{2,1}.
\end{eqnarray*}

This is an obstacle in constructing an eigenfunction defined
on the whole real axis.
The prescription is to demand that the coefficient of
 $y_{1}^{(-)}$ must vanish if $E$ is an eigenvalue.
Consequently, it is proportional to the spectral determinant.

The coefficient is essentially equal to
the (2,1) component of ${\cal M}_{0,2 m}^{(+)}$, and 
it reads in terms of the original $\phi$ function as 

$$
\frac{q^{\alpha(m+1)}}{2}
\begin{vmatrix}
\phi(x,\alpha, E),            & \phi(-x,\alpha, E) \\
\partial_x \phi(x,\alpha, E),   &  \partial_x  \phi(-x,\alpha,E)  \\
\end{vmatrix}.
$$

The $x$ dependencies are spurious as the entities are
products of Stokes multipliers which are obviously $x$ independent.
We  adopt the simplest choice  $x=0$.
The coefficient is now proportional to 
$\phi(0,\alpha,E) \partial_x \phi(x,\alpha,E)|_{x=0}$.
Thus we conclude that for an eigenvalue $E^{(+)}_j$ of ${\cal H}^{(+)}(x,\alpha)$,
\begin{equation}
\phi(0, \alpha, E^{(+)}_j) =0 \quad\text{ or  } \quad
 \partial_x \phi(x, \alpha, E^{(+)}_j)|_{x=0}=0 
\end{equation}
must hold.

We can repeat the same argument starting from 
${\cal H}^{(-)}(x,\alpha)$ on the positive real axis.

The above observation may lead to the identification
\begin{eqnarray}
\phi(0,\epsilon \alpha,E) &\sim& 
 D^{(\epsilon)}_-(E):= \prod_j (1- E^{(\epsilon)}_{-,j}),\\
\partial_x \phi(x,\epsilon \alpha,E)|_{x=0} &\sim& 
 D^{(\epsilon)}_+(E) := \prod_j (1- E^{(\epsilon)}_{+,j}) . 
\label{ident}
\end{eqnarray}

The lower sign signifies the parity:
the positive parity means a contribution from
symmetric wave function.
The product must be taken
over  eigenvalues with the corresponding parity.
The total set eigenvalues $\{ E^{(\epsilon)}_j   \}$  of 
${\cal H}^{(\epsilon)}(x,\alpha)$ 
consists of two subsets,
$\{ E^{(\epsilon)}_j   \}$ =$\{ E^{(\epsilon)}_{+,j}   \} \cup 
\{ E^{(\epsilon)}_{-,j}  \}   $ and
$D^{(\epsilon)}(E) = D^{(\epsilon)}_{+}(E) D^{(\epsilon)}_-(E)$.

%

We comment on the relation of the present result
to an existing solved model.
$T_{1,1}(E)$ in (\ref{pstransfer}) can be represented,
utilizing (\ref{ident}), as
\begin{equation}
T_{1,1}(E)=
q^{\alpha-1} \frac{D^{(+)}_-(E)}{D^{(+)}_-(q^4 E)}
+ 
q^{2 \alpha} \frac{D^{(+)}_-(E)D^{(-)}_-(q^6 E) }
      {D^{(+)}_-(q^4 E) D^{(-)}_-(q^2 E)}
+
q^{\alpha+1} \frac{D^{(-)}_-(q^6 E) }
      { D^{(-)}_-(q^2 E)},
\label{dvf11}
\end{equation}
where we safely choose  $x=0$ in the rhs. 

The above expression has similarity to
the dressed vacuum form (DVF) of the (unfused) transfer matrix for
the 3-state PS model with grading $(+,-,+)$.
The latter can be found in eq.(3.1) and (3.2) of \cite{Schulz83, JKSfusion}.
The spectral parameter $v$ corresponds to energy in the Schr{\"o}dinger operator,
precisely,  $E=\exp(\frac{2 \pi v}{M+1})$.

The spectral determinants have the following identification to 
the eigenvalues of Baxter's $Q$ operators,
\begin{eqnarray*}
D^{(+)}_-(E) &=& Q_2(v+\frac{j-2}{2}i)  \\
D^{(-)}_-(E) &=& Q_1(v+\frac{j-1}{2}i).  \\
\end{eqnarray*}
Those with positive parity may be identified with the
second solutions of  Baxter's $Q$ operators.

The scalar factors (vacuum expectation values)
$f_a(x), g_a(x)$ in  \cite{JKSfusion} 
depend on the choice of the quantum space.
We assume that the present quantum space space gives the simple
scalars as in (\ref{dvf11}).
In this sense, $T_{1,1}(E)$ exhibits 
the hidden $U_q(\widehat{gl}(2|1))$ symmetry
behind the present Schr{\"o}dinger operator,
just as in the $U_q(\widehat{sl}(2))$ symmetry for $\alpha=0$ problem.
This coincidence can be observed further.
 We have checked up to certain value of $k$
that the (1,1) element and the (2,2) element of ${\cal M}^{(+)}_{0,2 k}$
coincide with DVF of symmetric fusion transfer matrices
$\Lambda^{(1)}_{k}$ and $-\Lambda^{(1)}_{k-1}$ in
\cite{JKSfusion}, respectively.
The interpretation of the (1,2) and the (2,1) element, in terms of 
fusion transfer matrices, is still an open
problem.
%

One can adopt another description of $T_{1,1}$.
The (2,1) component of (\ref{single}) results
\begin{equation}
\tau^{(\epsilon)}_0 =\frac{\partial y_0^{(\epsilon)}}{  \partial y_1^{(-\epsilon)} }
+\frac{\partial y_2^{(\epsilon)} } {  \partial y_1^{(-\epsilon)} }  . 
\label{tqdash}
\end{equation}

Proceeding as above, we arrive at,

\begin{equation}
T_{1,1}(E)=
q^{\alpha+1} \frac{D^{(+)}_+(E)}{D^{(+)}_+(q^4 E)}
+ 
q^{2 \alpha} \frac{D^{(+)}_+(E)D^{(-)}_+(q^6 E) }
      {D^{(+)}_+(q^4 E) D^{(-)}_+(q^2 E)}
+
q^{\alpha-1} \frac{D^{(-)}_+(q^6 E) }
      { D^{(-)}_+(q^2 E)}.
\label{dvf11dash}
\end{equation}

In the next section, we determine the energy levels by utilizing the
above results.


\section{Nonlinear Integral equations for eigenvalue problem}

The Bethe ansatz equations follow from the pole-free
property of $T_{1,1}(E) $ on the real $E$ axis,
\begin{eqnarray}
q^{\alpha-1} \frac{D^{(-)}_+ (q^{2}E^{(+)}_{+,j}  )}
                  {D^{(-)}_+( q^{-2}E^{(+)}_{+,j}   )  }
&=& 
q^{-\alpha-1} \frac{D^{(+)}_+ (q^{2}E^{(-)}_{+,j}  )}
                  {D^{(+)}_+( q^{-2}E^{(-)}_{+,j}   )  }
=-1,   \\
q^{\alpha+1} \frac{D^{(-)}_- (q^{2}E^{(+)}_{-,j}  )}
                  {D^{(-)}_-( q^{-2}E^{(+)}_{-,j}   )  }
&=& 
q^{-\alpha+1} \frac{D^{(+)}_- (q^{2}E^{(-)}_{-,j}  )}
                  {D^{(+)}_-( q^{-2}E^{(-)}_{-,j}   )  }
=-1 .
\end{eqnarray}
To their analysis, we apply 
the strong machinery in solvable models, the
method of nonlinear integral equations.
Hereafter we shall confine ourselves to the case
$0\le \alpha \le M$ where energies are nonnegative.
The simple pattern of energy spectrum permits the 
following simple-mind choice of auxiliary functions,
\begin{eqnarray}
a^{(\epsilon)}_{\epsilon'}(E) &:=& q^{\epsilon \alpha -\epsilon'}
\frac{D^{(-\epsilon)}_{\epsilon'} (q^{2} E )}
                  {D^{(-\epsilon)}_{\epsilon'} ( q^{-2} E )  } 
  \label{smalla} \\
A^{(\epsilon)}_{\epsilon'}(E) &:=& 1+a^{(\epsilon)}_{\epsilon'}(E).  
\label{capitala}
\end{eqnarray}
Thus 
\begin{equation}
A^{(\epsilon)}_{\epsilon'}(E^{(\epsilon)}_{\epsilon', j})=0.
\end{equation}
Remember  that  $\epsilon (=\pm 1)$ represents the signature
of the perturbation while $\epsilon' (=\pm 1)$ denotes the parity.

In addition, we need  some inputs about the asymptotic behaviors 
from the  WKB method.
Fortunately, they are already available \cite{DT1} as the existence of
lower power term does not alter them.

\begin{eqnarray}
\ln D^{(\epsilon)}_{\pm}(E) &\sim& \frac{a_0}{2} (-E)^{\mu},
\qquad |E| \rightarrow \infty, |{\rm arg } (-E) | < \pi
\label{wkb1} \\
b_0 E_j^{(\epsilon)}  &\sim& 2 \pi (j+\frac{1}{2}), \quad j \rightarrow \infty 
\label{wkbapp}  \\
\mu&=& \frac{M+1}{2M}, \quad a_0=\frac{b_0}{2 \sin \mu \pi}, \nonumber \\
b_0 &=& \frac{\pi^{1/2} \Gamma(\frac{1}{2M})}{M \Gamma(\frac{3}{2}+\frac{1}{2M})}. 
\nonumber
\end{eqnarray}

There might be several routes to reach  nonlinear integral equations
among $a^{(\epsilon)}_{(\epsilon')}$ and
$A^{(\epsilon)}_{(\epsilon')}$.
Here we choose the quickest way \cite{DT1, DT3, DeVe92,BLZ2} 
which fully exploits the fact
that zeros of  $D^{(\epsilon)}_{\pm}(E)$ are
on the positive real $\theta$ axis.
In addition, we assume that there are no zeros of $A^{(\epsilon)}_{\pm}(E)$
inside the narrow strip including the positive real axis 
other than those from zeros of  $D^{(\epsilon)}_{\pm}(E)$.

Apparently we have
\begin{eqnarray*}
\log a^{(\epsilon)}_{\epsilon'}(E)  &=& 
 \frac{(\epsilon \alpha - \epsilon') \pi}{M+1} i
+ \sum_{j} F(\frac{E}{E^{(-\epsilon)}_{\epsilon',j}})  \\
F(E) &=& \log \frac{1- q^2 E}{1-q^{-2} E}.
\end{eqnarray*}

The above assumption allows the representation of  
the summation part by an integral over contour ${\cal C}_E$
which surrounds the positive real axis counterclockwise,

\begin{equation}
\log a^{(\epsilon)}_{\epsilon'}(E)  = 
\frac{(\epsilon \alpha -\epsilon') \pi}{M+1} i
+ \frac{1}{2\pi i} \int_{ {\cal C}_E } dE' F(\frac{E}{E'}) \partial_{E'}
 \log A^{(-\epsilon)}_{\epsilon'}(E') .
\label{nlie1}
\end{equation}

For convenience, we introduce a variable 
$\theta$ \cite{DT1} by
\begin{equation*}
E =  \exp(\theta/\mu) /\nu^2  \qquad
 \nu=(2M+2)^{-\frac{1}{2\mu}} /\Gamma(\frac{1}{2\mu})
\end{equation*}
which originates from the matching condition of the WKB result (\ref{wkb1})
and the $Q-$ operator analysis \cite{BLZ2, BLZ4}.

Let  $\mathfrak{a}^{(\epsilon)}_{\epsilon'} (\theta),
 \mathfrak{A}^{(\epsilon)}_{\epsilon'}(\theta)$ be
auxiliary functions defined in (\ref{smalla}), (\ref{capitala}) 
regarded as functions of $\theta$.
Then eq.(\ref{nlie1}) reads,
\begin{eqnarray*}
\log \mathfrak{a}^{(\epsilon)}_{\epsilon'} (\theta)  &=& 
\frac{(\epsilon \alpha -\epsilon') \pi}{M+1} i
+ \frac{1}{2\pi i} \int_{ {\cal C}_\theta } d\theta'  
G(\theta-\theta') \partial_{\theta'}
 \log \mathfrak{A}^{(-\epsilon)}_{\epsilon'}(\theta') \\
G(\theta) &=& \log \bigl( q^2 
 \frac{\sinh ( \frac{M \theta}{M+1}+i \frac{\pi}{M+1})  }
       {\sinh(\frac{M \theta}{M+1}-i \frac{\pi}{M+1})}                        
	                 \bigr )   \nonumber 
\end{eqnarray*}
where ${\cal C}_\theta$  encircles the whole real axis counterclockwise.

For the reason which will be supplemented,  we shall keep 
$\mathfrak{a}^{(\epsilon)}_{\epsilon'} (\theta)$
in the lower half plane but use  
$1/\mathfrak{a}^{(\epsilon)}_{\epsilon'} (\theta)$ in the upper half plane. 

This requirement  modifies the above expression as
\begin{eqnarray}
\log \mathfrak{a}^{(\epsilon)}_{\epsilon'} (\theta)  &=& 
\frac{(\epsilon \alpha -\epsilon') \pi}{M+1} i
- \frac{1}{2\pi i} 
  \int_{-\infty}^{\infty} \partial_{\theta} G(\theta-\theta'+ i 0) 
  \log \mathfrak{a}^{(-\epsilon)}_{\epsilon'} (\theta'-i 0)   \nonumber\\
& & + \frac{1}{\pi} \Im \Bigl( \int_{-\infty}^{\infty} d\theta'
 \partial_{\theta} G(\theta-\theta'+i 0) 
 \log \mathfrak{A}^{(-\epsilon)}_{\epsilon'}(\theta'-i 0)  \Bigr) \label{nlie2},
\end{eqnarray}
where  $\theta$ is assumed to possess small negative imaginary part.
The property  $({a}^{(\epsilon)}_{\epsilon'} (\theta) )^*
= \frac{1}{ {a}^{(\epsilon)}_{\epsilon'} (\theta^*)}$ is employed in the above
transformation.

We solve (\ref{nlie2}) in terms of 
$\log \mathfrak{a}^{(\epsilon)}_{\epsilon'} (\theta)$ to reach the final
expression of NLIE, 

\begin{eqnarray}
\ln \mathfrak{a}^{(\epsilon)}_{\epsilon'}(\theta) &=&
-\frac{i}{2} b_0 \nu^{-2 \mu} {\rm e}^{\theta} + 
\frac{\pi}{2} i (-\epsilon' + \epsilon \frac{\alpha}{M})   \nonumber\\
&+& 2 i \Im \{ \int_{-\infty}^{\infty} K_{1}(\theta-\theta'+i0)
         \ln \mathfrak{A}^{(\epsilon)}_{\epsilon'}(\theta'-i0) d\theta' \nonumber\\
& &   + \int_{-\infty}^{\infty} K_{2}(\theta-\theta'+i0)
                    \ln \mathfrak{A}^{(-\epsilon)}_{\epsilon'}(\theta'-i0) d\theta'
   \}.
\label{nlie}
\end{eqnarray}
The kernel functions read
\begin{eqnarray*}
K_1(\theta) &=& 
-\frac{1}{2 \pi} \int_{-\infty}^{\infty} e^{iw \theta} 
\frac{ \sinh^2 \frac{\pi (M-1) w}{2 M}}{\sinh\pi w \sinh \frac{\pi w}{M}}  \\
K_2(\theta) &=& 
-\frac{1}{2 \pi} \int_{-\infty}^{\infty} e^{iw \theta} 
\frac{\sinh \frac{\pi (M+1) w}{2 M} \sinh \frac{\pi (M-1)w}{2 M}  }
 {\sinh\pi w \sinh \frac{\pi w}{M}}.  \\
\end{eqnarray*}

Few remarks are in order.
\begin{enumerate}
\item As a consequence of connection rules, the integral equations are coupled
for auxiliary functions related to the positive and the negative coefficient
 of $x^{M-1}$ .
\item On the other hand,  equations with different parities are decoupled.
\item The constants are determined from the consistency by putting $\theta
\rightarrow -\infty$.  The $\alpha$ dependence is only summarized 
in these constants.
\item  The first term in the rhs is determined so that we recover the
result from the WKB method (\ref{wkb1}) 
by dropping contributions of integrals.
Clearly, $ \mathfrak{a}^{(\epsilon)}_{\epsilon'}(\theta)$ is  bounded 
in the upper-half plane.  This explains our choice of appropriate half planes
for auxiliary functions.

\end{enumerate}

The eigenvalues $\{E^{(\epsilon)}_{j,\pm} \}$ are evaluated by
$$
\ln \mathfrak{a}^{(\epsilon)}_{\pm}(\theta^{(\epsilon)}_{j,\pm})=(2 j+1) \pi i,
\quad \text{  and } \quad 
E^{(\epsilon)}_{j,\pm}=  \exp(\theta^{(\epsilon)}_{j,\pm}/\mu)/ \nu^2.
$$

More explicitly

\begin{eqnarray}
& & \frac{1}{2} b_0 \nu^{-2 \mu} {\rm e}^{\theta^{(\epsilon)}_{j,\epsilon'}}
=( 2 j+1- \epsilon' \frac{1}{2} + \epsilon \frac{\alpha}{2 M} ) \pi
           \nonumber \\
& & +2  \Im \{ 
   \int_{-\infty}^{\infty} K_{1}(\theta^{(\epsilon)}_{j,\epsilon'}-\theta'+i0)
              \ln \mathfrak{A}^{(\epsilon)}_{\epsilon'}(\theta') d\theta'
 + \int_{-\infty}^{\infty} K_{2}(\theta^{(\epsilon)}_{j,\epsilon'}-\theta'+i0)
                \ln \mathfrak{A}^{(-\epsilon)}_{\epsilon'}(\theta') d\theta'
           \}
           \nonumber \\
& &  \qquad(j \ge 0).  \label{detenergy}
\end{eqnarray}

We present examples of  numerical solutions
to (\ref{nlie2})  in Fig. 5.
The real and the imaginary parts of $\ln \mathfrak{A}^{(\pm)}_+$ are depicted
for $M=3, \alpha=1$.

\begin{equation*}
\boxed{\rm FIG. 5}
\end{equation*}

\section{benchmarks}
We shall check the nonlinear integral equations analytically for limiting cases
and numerically.

\noindent (1) $\alpha=0$ case \par
\noindent By putting,  
$\mathfrak{a}^{(+)}_{\pm}(E)=\mathfrak{a}^{(-)}_{\pm}(E)$
the coupled NLIE reduce to an identical integral equation.
Immediately seen, the result coincides with
the nonlinear integral equation in \cite{DT1}.
\par
\noindent (2) $\alpha=M$ case \par
\noindent In this case, we have a duality in energy spectra;
$ \{E^{(+)}_j \}$ coincide with $ \{E^{(-)}_j \}$ ,
except for $E^{(-)}_0=0$ in the latter.  
This degeneracy can be easily explained by the following representation
of the Hamiltonians \cite{facto, cann},
\begin{eqnarray*}
{\cal H}^{(-)}(x,\alpha=M) &=& {\cal D}^{\dagger} {\cal D}  \\
{\cal H}^{(+)}(x,\alpha=M) &=& {\cal D} {\cal D}^{\dagger} \\
{\cal D} &=& \frac{1}{i} \frac{d}{d x} -i x^M.
\end{eqnarray*}
Once an eigenvector 
${\cal H}^{(-)}(x,M) \psi^{(-)}_j =E^{(-)}_j \psi^{(-)}_j $
is found,
we can construct the eigenvector for ${\cal H}^{(+)}(x,\alpha)$
with the same energy by  $\psi^{(+)}_{j-1}:= {\cal D}  \psi^{(-)}_{j}$.
Only the exception is the $j=0$ case where ${\cal D}  \psi^{(-)}_{j=0}=0$.
It is interesting that the asymptotic form (\ref{asymptotics})
from  the WKB type argument
is exact for all $x$ in this case.
 \footnote
{I thank V.V. Bazhanov, R.J. Baxter and B. Nienhuis for pointing
out the explicit eigenfunction for  $\psi^{(-)}_0 $   right after 
my talk.}

The above facts can be also  verified from (\ref{nlie}).
Note that the rhs can be treated as the mod $2\pi i$ quantity.
Then the  choice $\alpha=M$ leads to the same coupled equations under 
identifications
${\mathfrak a}^{(+)}_+(\theta ) \leftrightarrow 
      {\mathfrak a}^{(-)}_-(\theta )$,
$ {\mathfrak a}^{(+)}_-(\theta ) \leftrightarrow
     {\mathfrak a}^{(-)}_+(\theta )$.
This explains the degeneracy of the spectra as it consists both
 from the negative and the positive parity contributions.
The zero energy case must be treated more separately.
By choosing $j=0, \epsilon=-\epsilon'=-1, \alpha=M$, 
we  find the first term in lhs of (\ref{detenergy})
is null.  So the first order approximation is 
$\theta^{(-)}_{+, 0}=-\infty$.  Actually this is exact as we determine
the constant terms so that NLIE are consistent in
$\theta \rightarrow -\infty$. See remark 3 after (\ref{nlie}).
This solution gives the missing energy 0.

Finally we present the preliminary numerical results for $M=3$.
%


\begin{table}[hbtp]
\begin{center}
\begin{tabular}{|c|c|c|c|c|c|c|}
\hline
$\alpha$&      IMSL  0-th&  IMSL 1st&    WKB  0th&   WKB 1st&  NLIE 0th& NLIE 1st\\
\hline
-2.5000&            0.22909& 2.3741&  $\diamond$&	2.36641 &  0.22872&	2.37175\\
-2.0000&            0.44007&  2.7962&       0*&	2.73228&        0.43969&	2.79688\\
-1.5000&            0.63726&  3.2028&        0.17736 &	3.09594&    0.63673&    3.20230\\
-1.0000&           0.81664&  3.5949&      0.38490&	3.45603&        0.81478& 3.59506\\
-0.50000&          0.98599&  3.9732&     0.59582 &  3.81142&   0.98547& 3.97303\\
0.0000&            1.1448&  4.3385 &       0.8008&  4.16123& 	1.1440&    4.3382\\
0.50000&           1.2943&  4.6917&      0.99516&  4.50476&  1.2931&  4.6918\\
1.0000&            1.4356&  5.0333&      1.1768& 4.84147&  1.43596&  5.0336\\
1.5000&            1.5696&  5.3642&     1.3456 &  5.17101&   1.57034&  5.3640\\
2.0000&            1.6972&  5.6850&     1.5024&  5.49313&   1.69667&  5.6842\\
2.5000&            1.8189&  5.9962&     1.6487&  5.80773&   1.81861&  5.9960\\
\hline
\end{tabular}
\end{center}
\end{table}

Table 1 shows the results from the  
IMSL package (dsleig.f), the (naive)  WKB method
and those obtained by solving the nonlinear equations.
The agreement is  not yet precise enough (typically 3-4 digits).
Some implement is still in need for the numerical accuracy.
Nevertheless, the NLIE data already 
show much improvement from the (naive) WKB results.

By the (naive) WKB method, we mean a self-consistent determination of
$E^{(\epsilon)}_j $ by
\begin{equation}
\oint |p| dx =  
\int_{-x_0}^{x_0}
  \sqrt{E^{(\epsilon)}_j-x^6-\epsilon \alpha x^2} dx =
(j+\frac{1}{2}) \pi
\label{wkbnaive}
\end{equation}
where  $E^{(\epsilon)}_j-x_0^6-\epsilon \alpha x_0^2=0$.
Particularly, for the value with asterisk, this method has subtlety.
Immediately seen, 
$E^{(\epsilon)}_0=0, x_0=2^{1/4}$ is a formal solution to (\ref{wkbnaive})
for $j=0, \epsilon=-, \alpha=2$.
It however involves an isolated turning point of the 2nd order at the origin 
if $E^{(\epsilon)}_0=0$, 
which spoils the simple application of the condition  (\ref{wkbnaive}).
The value with $\diamond$ has similar difficultly.
We  however skip further discussion on the validity on the (naive)  WKB method
as it is out of the present subject.

Summarizing, we check the consistency of (\ref{nlie}) in some limiting cases
and by numerical methods.

\section{Summary and Discussion}

In this report, the eigenvalue problem has been addressed for 
 the 1D quantum systems of which Hamiltonians include
double well potentials.
We have successfully derived the coupled NLIE which determine
 energy levels of the systems with potential terms of 
$\pm \alpha x^{M-1} + x^{2 M}$
at the same time.

The essence of our strategy is to utilize the following
 correspondences
between 1D quantum mechanics and  1+1 D solvable models,
\vskip 0.9cm
\begin{tabular}{rrl}
energy &    $\Longleftrightarrow$ &     spectral parameter \\
Stokes multipliers&  $\Longleftrightarrow$ &     transfer matrices \\
eigenfunctions or derivatives at $x=0$ &  $\Longleftrightarrow$ &    
    vacuum expectation values of $Q$ operators \\
\end{tabular}
\vskip 0.9cm
We are then entitled to apply the strong machinery of the latter
developed since  Baxter's revolution.

There are several open questions.
\begin{enumerate}
\item 
In this report we confine ourselves to the simplest
case $\alpha \le M$. 
 For $\alpha > M$, the existence of negative eigenvalues
ruins the analyticity assumptions on auxiliary functions.
Still, formal expressions of NLIE are possible which are
similar to   excited states TBA  equations.
The integration contour is, however,  not so simple as 
described here.  
This is an apparent drawback in actual numerical investigations.
The clever choice of auxiliary functions may be desired.
\item The understanding is lacking on the intrinsic reason 
why affine symmetry like
$U_q(\widehat{sl}(2) )$ or  $U_q(\widehat{gl}(2|1) )$ 
comes into play in
this simple 1D quantum mechanical model.
\item This is somewhat related to the above, but is the 
 most intriguing question. 
 {\bf Where is the Yang-Baxter equation
in the 1D Schr{\"o}dinger operator problem?}.
Once this is known, the fusion hierarchy, useful in the present study,
 is a mere corollary of it.
\end{enumerate}
We hope to answer these in the future publication.

\section*{Acknowledgments}
The author thanks V. V. Bazhanov, 
A. Kuniba, J.M. Maillard, B. Nienhuis P. Pearce, and C. Richard 
for comments, discussions and encouragement.
He also would like to thank the organizers of "Baxter's revolution
in mathematical physics".


\newpage
\begin{center}
{FIGURE CAPTIONS} 
\end{center}

\noindent Fig 1. An anharmonic oscillator perturbed by
a positive or a negative perturbation term.

\noindent Fig 2. The complex plane is divided into sectors.
${\cal S}_0$ and ${\cal S}_1$ are indicated as examples.

\noindent Fig 3. The FSS in 
${\cal S}_0$ and ${\cal S}_1$ are related by the matrix  
${\cal M}_{0,1}^{(+)}$.

\noindent Fig 4. The connection of FSS on the negative and the positive real axis
is accomplished by  ${\cal M}_{0,2 m}^{(+)}$.

\noindent Fig 5. Left: the real part of $\ln \mathfrak{A}^{(\pm)}_+$,
Right: the imaginary part of  $\ln \mathfrak{A}^{(\pm)}_+$.

\begin{center}
{TABLE CAPTIONS} 
\end{center}

\noindent Table 1. 
First two energy levels calculated by  the IMSL library, the (naive) 
WKB method and the NLIE method.  We choose $M=3$ and adopt various $\alpha$.
See asterisk and $\diamond$ for the text.

\newpage

\begin{figure}[hbtp]
\centering
  \includegraphics[width=10cm]{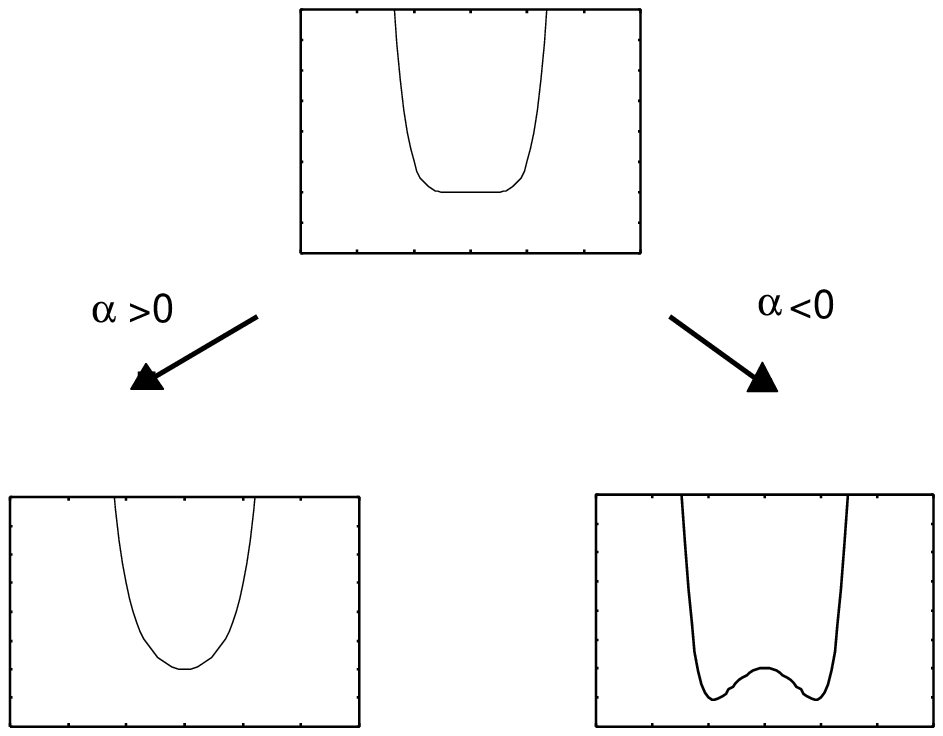}
\caption{}
\label{pot}
\end{figure}

\newpage
 \begin{figure}[hbtp]
\centering
  \includegraphics[width=10cm]{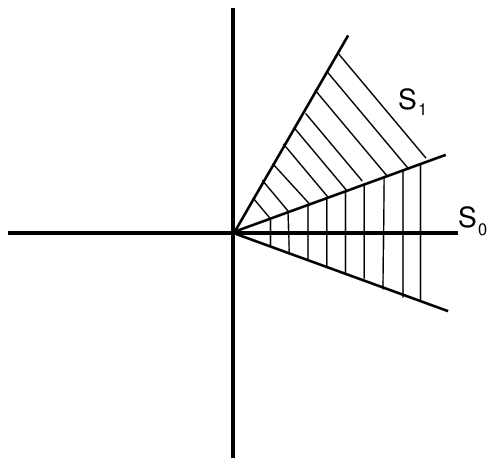}
\caption{}
\label{sec}
\end{figure}

\newpage
\begin{figure}[hbtp]
\centering
  \includegraphics[width=10cm]{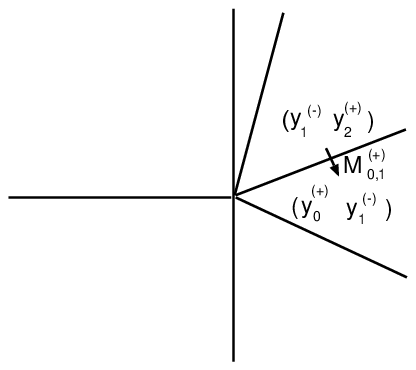}
\caption{}
\label{stokesm}
\end{figure}

\newpage
\begin{figure}[hbtp]
\centering
  \includegraphics[width=10cm]{realaxis.eps}
\caption{}
\label{connection}
\end{figure}

\newpage
 \begin{figure}[hbtp]
\centering
{  \includegraphics[width=5cm]{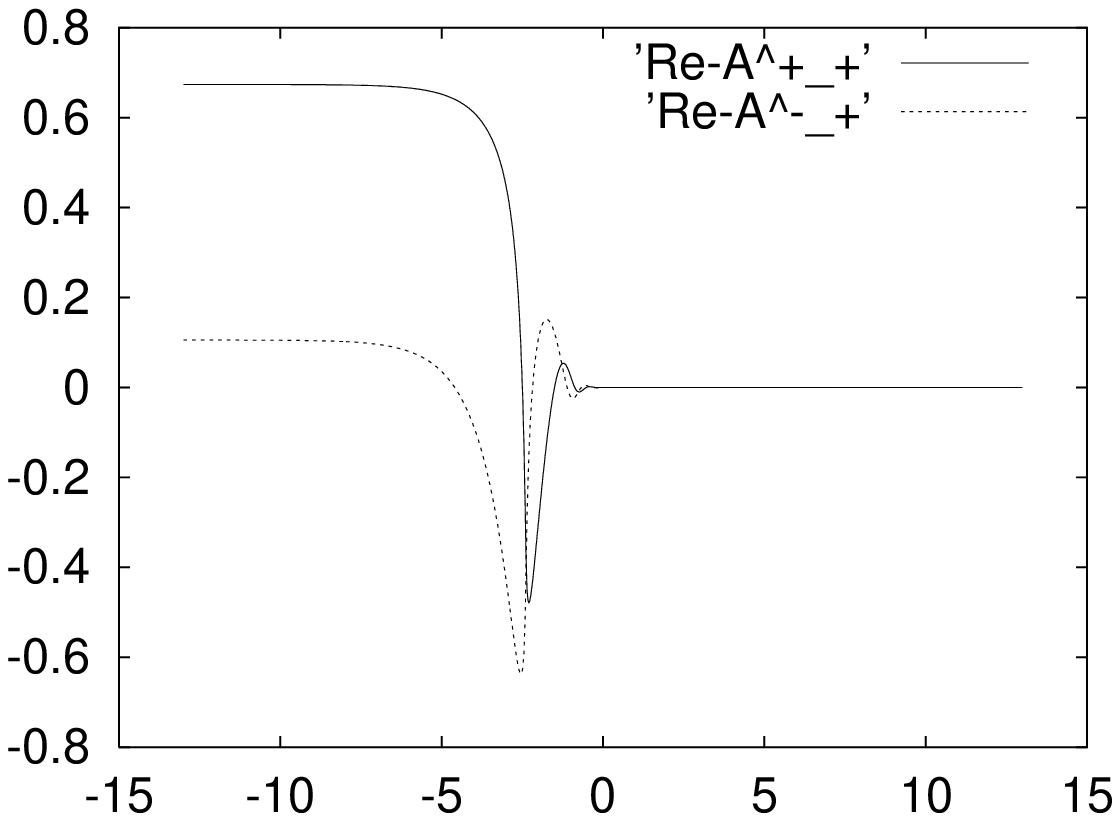} \hspace{1cm}
  \includegraphics[width=5cm]{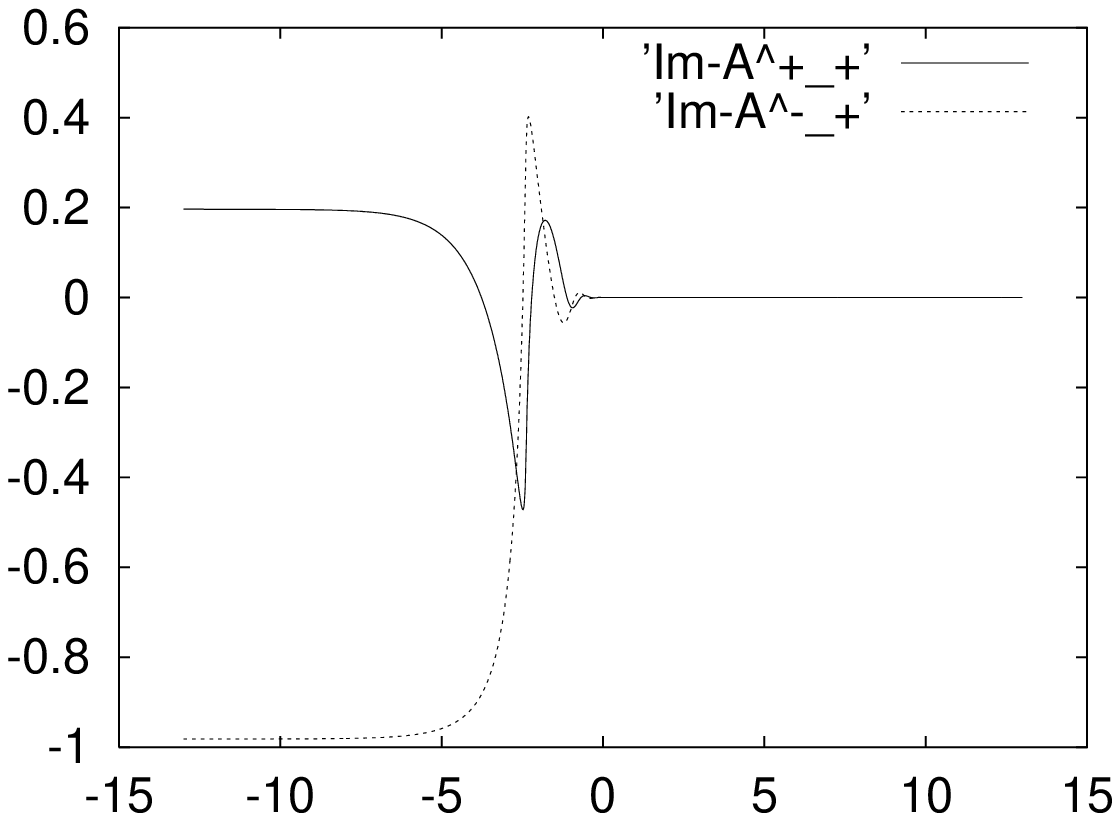} }
\caption{}
\label{afig}
\end{figure}

\end{document}